\documentclass[conference,a4paper]{IEEEtran}

\usepackage{cite}

\usepackage{makeidx,epsfig}
\usepackage{setspace,graphicx,multirow}

\usepackage{amsfonts,latexsym,amssymb,amsthm}

\usepackage{graphicx}
\usepackage{epstopdf}

\usepackage{caption3} 
\DeclareCaptionOption{parskip}[]{} 
\usepackage[small]{caption}

\usepackage{subcaption}

\usepackage{array}

\usepackage[cmex10]{amsmath}

\usepackage{url}

\usepackage{multirow}
\usepackage{hhline}

\usepackage{amsmath}
\usepackage{algorithm}
\usepackage[]{algpseudocode}
\usepackage{varwidth}

\makeatletter
\def\BState{\State\hskip-\ALG@thistlm}
\makeatother

\usepackage{algcompatible}

\usepackage{fixltx2e}

\newcommand*\diff{\mathop{}\!\mathrm{d}}

\usepackage{enumitem}

\makeatletter
  \newcommand\tinyv{\@setfontsize\tinyv{7pt}{9}}
\makeatother

\usepackage{authblk}

\title{Blocking Probability and Spatial Throughput Characterization for Cellular-Enabled UAV Network with Directional Antenna}
\author[]{Jiangbin Lyu}
\author[]{Rui Zhang}
\affil[]{\small Department of Electrical and Computer Engineering, National University of Singapore\\
email: \{elelujb, elezhang\}@nus.edu.sg}

\renewcommand\Authands{ and }

\begin{document}
\bibliographystyle{IEEEtran}
\bstctlcite{IEEEexample:BSTcontrol}

%

\maketitle

\begin{abstract}

The past few years have witnessed a tremendous increase on the use of unmanned aerial vehicles (UAVs) in civilian applications, which increasingly call for high-performance communication between UAVs and their ground clients, especially when they are densely deployed.  
To achieve this goal, cellular base stations (BSs) can be leveraged to provide a new and promising solution to support massive UAV communications simultaneously in a cost-effective way. 
However, different from terrestrial communication channels, UAV-to-BS channels are usually dominated by the light-of-sight (LoS) link, which aggravates the co-channel interference and renders the spatial frequency reuse in existing cellular networks ineffective.
In this paper, we consider the use of  a directional antenna at each UAV to confine the interference to/from other UAV users within a limited region and hence improve the spatial reuse of the spectrum.
Under this model, a UAV user may be temporarily blocked from communication if it cannot find any BS in its antenna main-lobe, or it finds that all BSs under its main-lobe are simultaneously covered by those of some other UAVs and hence suffer from strong co-channel interference.
Assuming independent homogeneous Poisson point processes (HPPPs) for the UAVs' and ground BSs' locations respectively, we first analytically derive a closed-form upper bound for the UAV blocking probability and then characterize the achievable average spatial throughput of the cellular-enabled UAV communication network, in terms of various key parameters including the BS/UAV densities as well as the UAV's flying altitude and antenna beamwidth.
Simulation results verify that the derived bound is practically tight, and further show that adaptively adjusting the UAV altitude and/or beamwidth with different BS/UAV densities can significantly reduce the UAV blocking probability and hence improve the network spatial throughput.
\end{abstract}
%

\section{Introduction}


With their enhanced functionality and ever-reducing cost, unmanned aerial vehicles (UAVs) have found fast-growing applications over recent years in the civilian domain such as for traffic control, precise agriculture, aerial imaging, search and rescue, and aerial communication platform, among others. 
As the number of UAVs and the demand for new UAV applications increase explosively in the near future, it is imperative to devise new solutions to support high-performance communications for UAVs, even when they are densely deployed.
However, at present, almost all UAVs rely on the simple direct point-to-point communication with their ground clients over the unlicensed spectrum (e.g., ISM 2.4GHz), which is typically of limited data rate, unreliable, insecure, vulnerable to interference, and can only operate within the visual line-of-sight (LoS) range. 

To enable massive UAV communications simultaneously, base stations (BSs) in the existing 4G (fourth-generation) LTE (Long Term Evolution) or forthcoming 5G (fifth-generation) cellular networks and beyond can be leveraged to provide a new and promising solution\cite{UAVinterferenceMagazine,SkyIsNotLimit}, thanks to their almost ubiquitous accessibility worldwide and superior performance.
As a result, cellular-enabled UAV communications are expected to achieve orders-of-magnitude performance improvement over the conventional point-to-point UAV-ground communications, in terms of all of reliability, security, coverage and throughput.

In fact, the 3rd Generation Partnership Project (3GPP) has recently started a new work item to discuss the various issues and their solutions for UAV communications using the current LTE BSs\cite{3GPPworkItemUAV}.
Moreover, there have been increasingly more field trials conducted on using terrestrial cellular networks to provide wireless connectivity for UAVs \cite{KDDItrials,QualCommDroneReport}.
Different from the terrestrial communication channels between ground user equipments (UEs) and BSs, it is reported in \cite{QualCommDroneReport} that the UAV-BS channels are usually dominated by the LoS link. On one hand, LoS channel does not suffer from multi-path fading or shadowing, and thus is more favorable than terrestrial channels from the perspective of each individual link between a UAV user and its communicating BS.
On the other hand, however, due to LoS channels, UAVs may generate more uplink interference to neighboring cells while also receiving more interference from them in the downlink. Specifically, there are two new tiers of interference in a cellular network supporting both ground UEs and UAV users, which are respectively the interference between UAV users and ground UEs, and that among different UAVs even when they are not densely distributed.
In this paper, we assume that a dedicated channel is assigned for the exclusive use by UAVs so that there is no interference between ground UEs and UAV users, and henceforth we focus our study on dealing with the interference issue among UAVs.
Note that in practice the assigned channel for UAVs can be opportunistically reused by ground UEs when there are no UAV users in the vicinity.

\begin{figure}
\centering
   \includegraphics[width=1\linewidth]{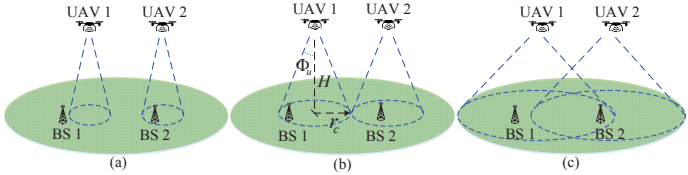}
\caption{Cellular-enabled UAV communication with directional antenna.\vspace{-2ex}}\label{Schematic}
\end{figure}

We consider the use of a directional antenna at each UAV that beams downward to communicate with its associated ground BS in the antenna main-lobe. This can effectively confine the interference to/from other UAVs within a limited region and hence significantly improve the spatial reuse of the assigned channel for multi-UAV communications.
As shown in Fig. \ref{Schematic}, each UAV has a corresponding coverage region on the ground plane given its antenna beamwidth, within which the UAV can associate with a BS (if any) for communication. 
If there is no BS in its coverage region (e.g., see UAV 1 in Fig. \ref{Schematic}(a)) or all the BSs therein are simultaneously covered by other UAVs (thus, suffering from strong co-channel interference; see, e.g., UAV 2 in Fig. 1(c)), then the UAV is said to be temporarily \textit{blocked} from communication. 
Given a fixed UAV altitude, smaller UAV beamwidth corresponds to larger antenna gain but smaller coverage region on the ground, thus increasing the UAV blocking probability.
In contrast, larger UAV beamwidth corresponds to smaller antenna gain but larger coverage region, which increases the chance of covering more BSs (thus helps reducing the UAV blocking probability on one hand), but also leads to more overlaps with the coverage regions of other UAVs (thus results in increased UAV blocking probability on the other hand).      
Therefore, there exists an optimal beamwidth to minimize the UAV blocking probability; in other words, under the optimal beamwidth, the probability that each UAV has at least one \textit{exclusively covered} BS in its coverage region (e.g., see Fig. \ref{Schematic}(b)) is maximized.
Similar trade-offs exist for the UAV altitude control in minimizing the UAV blocking probability given a fixed UAV beamwidth.
Therefore, both the UAV altitude and beamwidth affect the direct UAV-BS channel gain and UAV blocking probability, and hence it is worth investigating their effects on the achievable average spatial throughput of all UAV users in the considered cellular-enabled UAV network.

To this end, we model the BS and UAV locations as two independent homogeneous Poisson point processes (HPPPs) with given densities.
First, we derive a closed-form upper bound for the UAV blocking probability. Based on this result, we then characterize the achievable average spatial throughput of the network, in terms of various key parameters including the BS/UAV densities as well as the UAV's flying altitude and antenna beamwidth.
Simulation results verify that the derived bound is practically tight, and further show that adaptively adjusting the UAV altitude and/or beamwidth with different BS/UAV densities can significantly reduce the UAV blocking probability and hence improve the network spatial throughput.

Besides cellular-enabled UAV communications considered in this paper, it is worth noting that there has been recently another line of research in UAV communications, which aims to employ UAVs as aerial platforms to provide wireless communication service to ground users in scenarios where there are insufficient or even unavailable ground BSs\cite{ZengUAVmag}, such as UAV-enabled ubiquitous coverage or drone small cells (DSCs) \cite{CyclicalLyu,PlacementLyu,UAVGBSglobecom,DroneSmallCell,3Dplacement,ZhangWeiSmallCellJSAC,UAVoverload,UAVpublicSafety,WuQingQingUAVTWC}, UAV-enabled mobile relaying \cite{UAVrelay,ZengMobileRelay} and UAV-enabled information dissemination/data collection \cite{ZhanChengDataCollection}, etc. 

%

\textit{Notations}: $\mathbb{R}$ denotes the set of real numbers; $\mathbb{E}[\cdot]$ denotes the expectation of a random variable; $\|\cdot\|$ denotes the Euclidean norm; $|\cdot|$ takes the cardinality of a set; $\setminus\cdot$ denotes the set minus operation; $\cup$ denotes the set union; $\cap$ denotes the set intersection; and $\emptyset$ denotes the empty set.

\section{System Model}\label{SectionModel}
We consider uplink transmission from UAVs to ground BSs, whereas the results can be similarly applied to the downlink transmission.
The BS locations are modeled by a 2-dimensional (2D) HPPP $\Lambda_b$ on the ground plane with given density $\lambda_b$ BSs per square meter (BSs/m$^2$).
Denote the set of BS locations as $\mathcal{W}\triangleq\{\bold w_k\in\mathbb{R}^2|k\in\Lambda_b\}$, where $\bold w_k$ is the 2D coordinate of a BS $k\in\Lambda_b$.
For the purpose of exposition, we assume that the UAVs fly at the same altitude $H$, and follow a 2D HPPP $\Lambda_u$ (independent from $\Lambda_b$) in the horizontal plane with given density $\lambda_u$ UAVs/m$^2$.
Denote the set of UAV horizontal locations as $\mathcal{U}\triangleq\{\bold u_m\in\mathbb{R}^2|m\in\Lambda_u\}$, where $\bold u_m$ is the 2D coordinate of a UAV $m\in\Lambda_u$ projected on the ground plane. In the case with moving UAVs, $\mathcal{U}$ models a snapshot of the UAV network.

\begin{figure}
\centering
   \includegraphics[width=0.86\linewidth,  trim=240 40 20 30,clip]{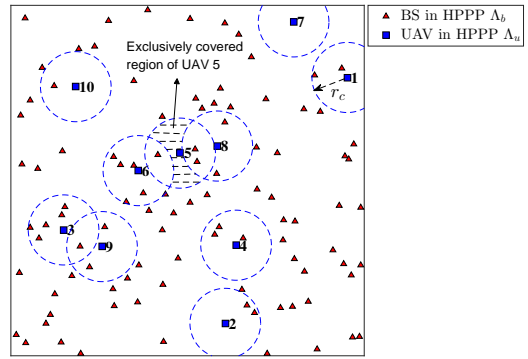}
\caption{Illustration of UAV blocking scenarios.\vspace{-2ex}}\label{PHP}
\end{figure}

\subsection{Channel Model}

We assume that each UAV is equipped with a directional antenna pointing downward towards the ground, whose azimuth and elevation half-power beamwidths are both $2\Phi_u$ radians (rad) with $\Phi_u\in(0,\frac{\pi}{2})$, as shown in Fig. \ref{Schematic}(b). Furthermore, the corresponding antenna gain in direction $(\phi,\varphi)$ can be practically approximated as
\begin{align}\label{UAVantenna}
G_u(\phi,\varphi)=
\begin{cases}
G_0/\Phi_u^2, & \textrm{$-\Phi_u\leq \phi\leq \Phi_u$, $-\Phi_u\leq \varphi\leq \Phi_u$;}\\
g_0\approx 0, & \ \textrm{otherwise,}
\end{cases}
\end{align}%
where $G_0=\frac{30000}{2^2}\times(\frac{\pi}{180})^2\approx 2.2846$; $\phi$ and $\varphi$ denote the azimuth and elevation angles, respectively \cite{balanis2016antenna}. Note that in practice, $g_0$ satisfies $0<g_0 \ll G_0/\Phi_u^2$, and for simplicity we assume $g_0= 0$ in this paper.
On the other hand, we assume for simplicity that each BS is equipped with an isotropic antenna of unit gain\footnote{In practice, the BS antenna is usually tilted downward to support ground UEs \cite{QualCommDroneReport}, and hence likely to communicate with UAVs in its sidelobes, which are assumed here to have uniform and unit antenna gain for simplicity.}.
Thus, the disk region centered at the UAV's projection on the ground with radius $r_c=H\tan\Phi_u$ corresponds to the ground \textit{coverage region} by the antenna main-lobe of the UAV. 
Consider a typical UAV 0 whose projection on the ground is at the origin.
The antenna gain of UAV 0 at altitude $H$ as seen by a ground BS at location $\bold w$ can be expressed as
\begin{align}\label{UAVantennaBS0}
G_{u,0}(\bold w,H)=
\begin{cases}
G_0/\Phi_u^2, & \textrm{$\|\bold w\|\leq H\tan\Phi_u$;}\\
0, & \ \textrm{otherwise.}
\end{cases}
\end{align}%

The UAV-BS channel is usually dominated by the LoS link, as justified by Qualcomm's trial report \cite{QualCommDroneReport}. The received channel power gain from UAV 0 to a BS at $\bold w$ thus follows the free-space path loss model given by
\begin{equation}
h(\bold w,H)=\kappa_0 d^{-2}(\bold w,H)=\frac{\kappa_0}{\|\bold w\|^2+H^2},
\end{equation}
where $\kappa_0=(\frac{4\pi f_c}{c})^{-2}$ denotes the channel power gain at a reference distance of 1 meter (m), with $f_c$ denoting the carrier frequency and $c$ denoting the speed of light; and $d(\bold w,H)=\sqrt{\|\bold w\|^2+H^2}$ is the UAV-BS link distance.

\subsection{UAV to BS Association}

Each UAV aims to find a BS within its ground coverage region and associate with it for communication. If the UAV cannot find any BS within its coverage region, or all the BSs therein are simultaneously covered by other UAVs (thus potentially suffering from strong co-channel interference), then the UAV is said to be temporarily \textit{blocked}. The typical blocking cases are illustrated in Fig. \ref{PHP}.
The BSs in the HPPP $\Lambda_b$ are represented by triangles, and the UAVs in the independent HPPP $\Lambda_u$ are represented by squares, while the region within each dashed circle of radius $r_c$ represents the coverage region of the corresponding UAV. 
There are two blocking cases for UAVs. First, a UAV may find no BSs inside its coverage region, e.g., UAV 7 in Fig. \ref{PHP}. Second, a UAV may have BSs inside its coverage region but all of them are also covered and hence interfered with by other UAVs, e.g., UAV 5 in Fig. \ref{PHP}, which has no BSs inside its exclusively covered region (the shadowed region).
In other words, a UAV is blocked if it cannot find a BS in its \textit{exclusively covered region} which is non-overlapping with the coverage region of any other UAV.


In practice, a UAV might send association requests to the BSs within its coverage region, and associate with one of the BSs with desirable signal-to-interference-plus-noise ratio (SINR).
As a result, a UAV will choose to associate with one of its exclusively covered BSs, if any.
Otherwise, the UAV-BS link suffers from low SINR and is hence assumed to be blocked, either due to strong interference from other UAVs, or because the nearest BS is outside its antenna main-lobe.
Note that time-/frequency-division multiple access (TDMA/FDMA) schemes could be applied to orthogonalize the transmissions of two or more interfering UAV-BS links.
However, such schemes require network coordination among multiple BSs and are thus difficult to implement in practice, especially in the case where the UAVs are moving at high speed.
Therefore, in this paper we only consider the worst-case performance where a UAV-BS link is blocked when it suffers from strong co-channel interference.

\subsection{UAV Blocking Probability}\label{SectionBlock}

The blocking probability of a UAV in the HPPP $\Lambda_u$ is defined as the probability that
the UAV cannot find any BS in the independent HPPP $\Lambda_b$ inside its exclusively covered region.
Specifically, the coverage region of a UAV $m\in\Lambda_u$ is the disk region of radius $r_c=H\tan\Phi_u$ centered at UAV $m$'s ground projection $\bold u_m$, denoted as $\mathcal{B}(\bold u_m,r_c)$. 
The exclusively covered region (or non-overlapping coverage region) of the typical UAV $0\in\Lambda_u$ can then be denoted as $\mathcal{A}_0\triangleq\mathcal{B}(\bold u_0,r_c) \setminus\bigcup\limits_{m\neq 0, m\in \Lambda_u}\mathcal{B}(\bold u_m,r_c)$.
Therefore, the UAV blocking probability is defined as
\begin{equation}\label{PBdefinition}
\textrm{P}_{\textrm{B}}\triangleq\textrm{Pr}\big\{\mathcal{W}\cap \mathcal{A}_0=\emptyset\big\}.
\end{equation}
Denote $A$ as the area of $\mathcal{A}_0$.
Since the HPPP $\Lambda_b$ and HPPP $\Lambda_u$ are independent, 
the probability that no BS in $\Lambda_b$ lies in the region $\mathcal{A}_0$ is given by $\exp(-\lambda_b A)$. The UAV blocking probability is then given by
\begin{equation}\label{PBintegral}
\textrm{P}_{\textrm{B}}=\int_0^{S}\exp(-\lambda_b A)f(A)\diff A,
\end{equation}
where $S=\pi r_c^2$ is the area of a full coverage disk of radius $r_c$, and $f(A)$ is the probability density function (p.d.f.) of the non-overlapping coverage area $A$.



The UAV blocking probability is thus a function of the coverage radius $r_c$ as well as the BS density $\lambda_b$ and UAV density $\lambda_u$, denoted as $\textrm{P}_{\textrm{B}}(\lambda_b,\lambda_u,r_c)$. Since $r_c=H\tan\Phi_u$, it is also a function of the UAV altitude $H$ and beamwidth $\Phi_u$, thus equivalently denoted as $\textrm{P}_{\textrm{B}}(\lambda_b,\lambda_u,H,\Phi_u)$.

\subsection{Link Data Rate and Average Spatial Throughput}

Assume that each UAV has transmit power $P$ in Watt.
Suppose that the typical UAV user 0 is associated with an exclusively covered BS 0 at location $\bold w_0$ where $\|\bold w_0\|\leq r_c=H\tan\Phi_u$, whose maximum achievable rate in bits/second/Hertz (bps/Hz) normalized to the channel bandwidth $W$ is given by
\begin{align}
R&=\log_2\bigg(1+\frac{P h(\bold w_0,H) G_{u,0}(\bold w_0,H)}{N_0 W}\bigg)\notag\\ 
             &=\log_2\bigg(1+\frac{\kappa_0 G_0 P}{\sigma^2 \Phi_u^2\big(\|\bold w_0\|^2+H^2\big)}\bigg)\label{RateUAV}\\
             &\geq \log_2\bigg(1+\frac{\kappa_0 G_0 P}{\sigma^2 \Phi_u^2\big((H\tan\Phi_u)^2+H^2\big)}\bigg)\label{eqEdge}\\
             &=\log_2\bigg(1+\frac{\kappa_0 G_0 P \cos^2\Phi_u}{H^2\sigma^2 \Phi_u^2}\bigg)\triangleq \bar R(H, \Phi_u),\label{RateUAVmin}
\end{align}
where the receiver noise is assumed to be additive white Gaussian noise (AWGN) with power spectrum density $N_0$ in Watt/Hz and $\sigma^2\triangleq N_0 W$ is the noise power over the bandwidth $W$.
For simplicity, we assume that each UAV transmits at a lower bound of its achievable rate, denoted by $\bar R$ in \eqref{RateUAVmin}, which corresponds to the BS at the edge of the UAV's coverage region (see \eqref{eqEdge}) and is a decreasing function of both $H$ and $\Phi_u$.
The achievable average spatial throughput of all UAV users in bps/Hz/m$^2$ is thus given by
\begin{equation}\label{throughput}
\theta(\lambda_b,\lambda_u,H, \Phi_u)\triangleq \lambda_u\big(1-  \textrm{P}_{\textrm{B}}(\lambda_b,\lambda_u,H, \Phi_u)\big) \bar R(H, \Phi_u).
\end{equation}


\section{Characterization of UAV Blocking Probability and Spatial Throughput}\label{SectionPb}

To the best of our knowledge, there is no further analytical expression of the UAV blocking probability $\textrm{P}_{\textrm{B}}$ given in \eqref{PBintegral} in the literature. In this section, we first derive a closed-form upper bound for $\textrm{P}_{\textrm{B}}$ and then characterize the achievable spatial throughput of the cellular-enabled UAV network defined in \eqref{throughput}.

\subsection{Upper Bound for $\textrm{P}_{\textrm{B}}$}\label{SectionPBcharacterization}
\begin{figure}
\centering
   \includegraphics[width=0.95\linewidth]{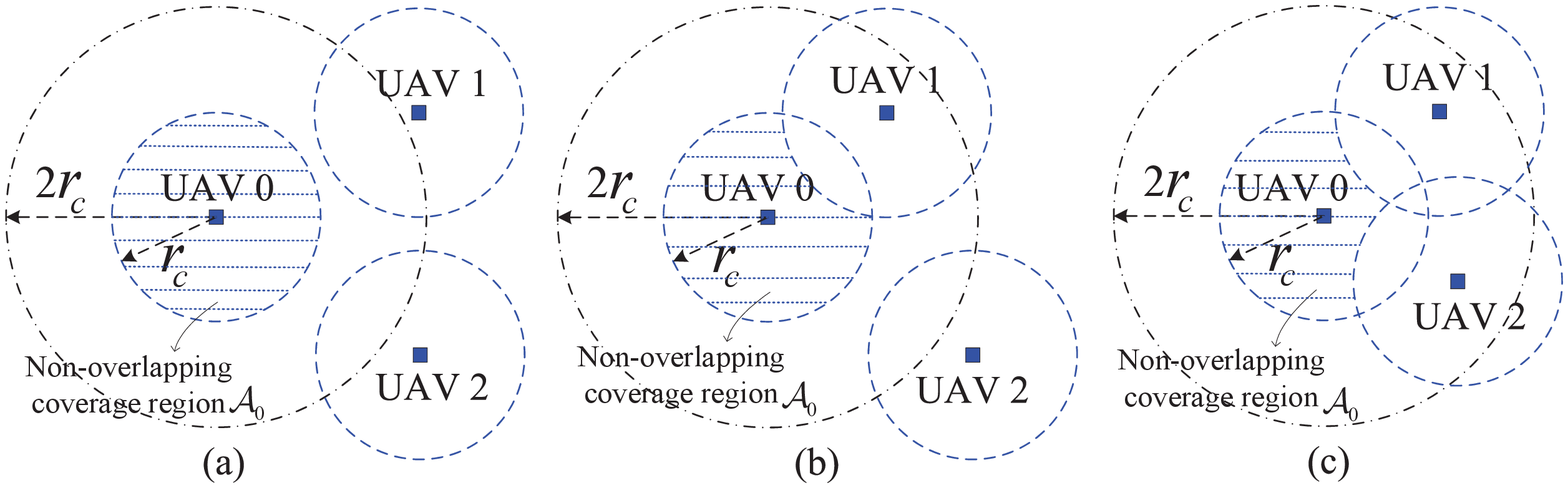}
\caption{Illustration of the non-overlapping coverage region $\mathcal{A}_0$.\vspace{-2ex}}\label{Cellular4UAVa0}
\end{figure}

According to \eqref{PBdefinition} and \eqref{PBintegral},
the major challenge in deriving $\textrm{P}_{\textrm{B}}$ lies in how to characterize the non-overlapping coverage region $\mathcal{A}_0$ and its area distribution $f(A)$.
For illustration, consider the typical UAV 0 in Fig. \ref{Cellular4UAVa0}.
If there is no UAV within distance $2r_c$ from UAV 0, then $\mathcal{A}_0$ is the full disk region of radius $r_c$ which is not overlapped by other UAVs, as shown in Fig. \ref{Cellular4UAVa0} (a).
If there is one or more UAVs within distance $2r_c$ from UAV 0, then $\mathcal{A}_0$ is the residue region of the coverage disk after carving out the overlapping parts (e.g., see Fig. \ref{Cellular4UAVa0} (b) and (c) where there are 1 and 2 overlapping UAVs, respectively).
As more UAVs overlap with UAV 0, the non-overlapping coverage area $A$ becomes more difficult to characterize.

Following the above illustration,
the blocking event of the typical UAV 0 can be decomposed into two sub-events: (1) A subset of UAVs $\mathcal{M}\subseteq \Lambda_u\setminus \{0\}$ have overlapping coverage region with UAV 0, i.e., $\bold u_m\in \mathcal{B}(\bold u_0,2r_c), \forall m\in \mathcal{M}$; (2) Conditioned on event (1), there is no BS in the non-overlapping coverage region $\mathcal{A}_0$ of UAV 0. 
Therefore, the UAV blocking probability in \eqref{PBdefinition} can be decomposed as
\begin{equation}\label{Pout2terms}
\textrm{P}_{\textrm{B}}=\sum\limits_{l=0}^{\infty}\textrm{Pr}\big\{|\mathcal{M}|=l\big\}\textrm{Pr}\big\{\mathcal{W}\cap \mathcal{A}_0=\emptyset\big | |\mathcal{M}|=l\big\}.
\end{equation}
In above, the first term $\textrm{Pr}\big\{|\mathcal{M}|=l\big\}$ is the probability that $l$ ($l\geq 0$) UAVs in the HPPP $\Lambda_u$ are within a disk of radius $2r_c$, which is expressed by following the Poisson distribution as
\begin{equation}\label{term1}
\textrm{Pr}\big\{|\mathcal{M}|=l\big\}=\frac{\big(4\pi r_c^2\lambda_u\big)^l \exp\big(-4\pi r_c^2\lambda_u\big)}{l!}, l=0,1,\cdots.
\end{equation}
The difficulty lies in characterizing the second term $\textrm{Pr}\big\{\mathcal{W}\cap \mathcal{A}_0=\emptyset\big | |\mathcal{M}|=l\big\}$ in \eqref{Pout2terms} with $l\geq 1$, which is the probability that no BS in the HPPP $\Lambda_b$ is inside the non-overlapping coverage region $\mathcal{A}_0$ under the condition that $|\mathcal{M}|=l\geq 1$ other UAVs have overlapping coverage region with UAV 0.

For the tractability of analysis, we first upper-bound the infinite series expansion of $\textrm{P}_{\textrm{B}}$ in \eqref{Pout2terms} by an expansion of finite terms. This is motivated by observing that the area $A$ of the non-overlapping coverage region $\mathcal{A}_0$ goes to zero as the number of overlapping UAVs $l$ increases, which makes $\textrm{Pr}\big\{\mathcal{W}\cap \mathcal{A}_0=\emptyset\big | |\mathcal{M}|=l\big\}\rightarrow 1$ as $l\rightarrow \infty$. Therefore, $\textrm{P}_{\textrm{B}}$ is upper-bounded by

\begin{footnotesize}
\vspace{-7pt}
\begin{align}\label{Pout2termsUpper}
\textrm{P}_{\textrm{B}}&\leq \sum\limits_{l=0}^{L}\textrm{Pr}\big\{|\mathcal{M}|=l\big\}\textrm{Pr}\big\{\mathcal{W}\cap \mathcal{A}_0=\emptyset\big | |\mathcal{M}|=l\big\}+\sum\limits_{l=L+1}^{\infty}\textrm{Pr}\big\{|\mathcal{M}|=l\big\}\notag\\
&=\sum\limits_{l=0}^{L}\textrm{Pr}\big\{|\mathcal{M}|=l\big\}\textrm{Pr}\big\{\mathcal{W}\cap \mathcal{A}_0=\emptyset\big | |\mathcal{M}|=l\big\}+1-\sum\limits_{l=0}^{L}\textrm{Pr}\big\{|\mathcal{M}|=l\big\}\notag\\
&=1-\sum\limits_{l=0}^{L}\textrm{Pr}\big\{|\mathcal{M}|=l\big\}\bigg(1-\textrm{Pr}\big\{\mathcal{W}\cap \mathcal{A}_0=\emptyset\big | |\mathcal{M}|=l\big\}\bigg),
\end{align}
\end{footnotesize}where $L\geq 1$ is an integer which is usually taken to be sufficiently large to achieve a tight upper bound.

Next, we focus on characterizing the probability term $\textrm{Pr}\big\{\mathcal{W}\cap \mathcal{A}_0=\emptyset\big | |\mathcal{M}|=l\big\}$.
Similar to \eqref{PBintegral}, we have
\begin{align}
\textrm{Pr}&\big\{\mathcal{W}\cap \mathcal{A}_0=\emptyset\big | |\mathcal{M}|=l\big\}\notag\\
&=\int_{A=0}^{S} \exp(-\lambda_b A) f\big(A\big | |\mathcal{M}|=l\big)\diff A \label{integral1}\\
&=\int_{\eta=0}^{1} \exp(-\lambda_b S\eta) f\big(\eta\big | |\mathcal{M}|=l\big)\diff \eta, \label{integral2}
\end{align}
where $\eta\triangleq A/S$ is the ratio of the non-overlapping coverage area $A$ to the full coverage disk area $S=\pi r_c^2$; $f\big(A\big | |\mathcal{M}|=l\big)$ and $f\big(\eta\big | |\mathcal{M}|=l\big)$ are the p.d.f. of $A$ and $\eta$ conditioned on $|\mathcal{M}|=l$, respectively.
It is generally difficult to obtain closed-form expressions for $f\big(\eta\big | |\mathcal{M}|=l\big)$ and hence $\textrm{Pr}\big\{\mathcal{W}\cap \mathcal{A}_0=\emptyset\big | |\mathcal{M}|=l\big\}$, since the overlapping region of UAV 0 and $l$ randomly located nearby UAVs can have an arbitrary shape.

Fortunately, we find a connection between $\textrm{Pr}\big\{\mathcal{W}\cap \mathcal{A}_0=\emptyset\big | |\mathcal{M}|=l\big\}$ in \eqref{integral2} and the moment generating function (m.g.f.)\footnote{The m.g.f. of a random variable $x$ is defined as $M_x(t)\triangleq\mathbb{E}[\exp(tx)]=\int_{-\infty}^{\infty} \exp(tx) f(x) \diff x, t\in\mathbb{R}$, where $f(x)$ is the p.d.f. of $x$.} of the random variable $\eta$ conditioned on $|\mathcal{M}|=l$. 
Specifically, denote the m.g.f. of $\eta$ conditioned on $|\mathcal{M}|=l$ as $M_{\eta,l}(t)\triangleq \mathbb{E}\big[\exp(t\eta)\big | |\mathcal{M}|=l\big]$, then we have
\begin{equation}\label{MGF}
\textrm{Pr}\big\{\mathcal{W}\cap \mathcal{A}_0=\emptyset\big | |\mathcal{M}|=l\big\}=M_{\eta,l}(-\lambda_b S).
\end{equation}
Based on the following theorem on m.g.f., we can therefore obtain an upper bound for $\textrm{Pr}\big\{\mathcal{W}\cap \mathcal{A}_0=\emptyset\big | |\mathcal{M}|=l\big\}$.

\newtheorem{theorem}{Theorem}
\begin{theorem}[D. Brook \cite{DBrook}]\label{theo1}
For a non-negative, real-valued random variable $x$, its m.g.f. $M_x(t)=\mathbb{E}[\exp(tx)]$ is upper-bounded by
\begin{equation}
M_x(t)\leq K_x(t)\triangleq 1-\alpha^2/\beta+(\alpha^2/\beta)\exp(\beta t/\alpha), \quad t\leq 0,
\end{equation}
where $\mathbb{E}[x]=\alpha$ and $\mathbb{E}[x^2]=\beta$ are the first and second moment of $x$, respectively.
\end{theorem}

Denote the first and second moment of $\eta$ conditioned on $|\mathcal{M}|=l$ as $\alpha_l\triangleq \mathbb{E}\big[\eta\big | |\mathcal{M}|=l\big]$ and $\beta_l\triangleq \mathbb{E}\big[\eta^2\big | |\mathcal{M}|=l\big]$, respectively.
Based on Theorem \ref{theo1}, we thus have
\begin{align}
&\textrm{Pr}\big\{\mathcal{W}\cap \mathcal{A}_0=\emptyset\big | |\mathcal{M}|=l\big\}=M_{\eta,l}(-\lambda_b S)\notag\\
&\leq 1-\alpha_l^2/\beta_l+(\alpha_l^2/\beta_l)\exp(-\lambda_b S \beta_l/\alpha_l)\triangleq K_{\eta,l}(-\lambda_b S).\label{Ketal}
\end{align}
Therefore, when $\alpha_l$'s and $\beta_l$'s are known for $l=0,1,\cdots,L$ where $L$ is a given positive integer, we can obtain a closed-form upper bound for the UAV blocking probability by substituting \eqref{term1} and \eqref{Ketal} into \eqref{Pout2termsUpper}, which yields
\begin{align}\label{LyuUpperBound}
&\textrm{P}_{\textrm{B}}^{\textrm{upper}}(\lambda_b,\lambda_u,r_c)\notag\\
&\triangleq 1-\sum\limits_{l=0}^{L} \frac{(4\lambda_u S)^l \exp(-4\lambda_u S)}{l!} \big(1-K_{\eta,l}(-\lambda_b S)\big),
\end{align}
where $S=\pi r_c^2$.
The method to obtain or estimate $\alpha_l$'s and $\beta_l$'s for $l\geq 0$ is given in Appendix \ref{appendixA}.

\subsection{Lower Bound for Spatial Throughput}

Based on the upper bound for the UAV blocking probability in \eqref{LyuUpperBound}, we can obtain a lower bound for the spatial throughput defined in \eqref{throughput}.
Since $r_c=H\tan\Phi_u$, we can also denote $\textrm{P}_{\textrm{B}}^{\textrm{upper}}(\lambda_b,\lambda_u,r_c)$ in \eqref{LyuUpperBound} as $\textrm{P}_{\textrm{B}}^{\textrm{uppper}}(\lambda_b,\lambda_u,H,\Phi_u)$. A lower bound for the achievable spatial throughput is then given by
\begin{equation}\label{throughputLower}
\theta^{\textrm{lower}}(\lambda_b,\lambda_u,H, \Phi_u)\triangleq \lambda_u\big(1-  \textrm{P}_{\textrm{B}}^{\textrm{uppper}}(\lambda_b,\lambda_u,H,\Phi_u)\big) \bar R(H, \Phi_u).
\end{equation}

\begin{table*}[t]\tinyv
\centering
\caption{1st and 2nd moment of the non-overlapping ratio $\eta$ in the coverage region of UAV 0 given $l$ overlapping UAVs.}
\addtolength{\tabcolsep}{-5.5pt}
\renewcommand{\arraystretch}{1.1}
\begin{tabular}{|c|c|c|c|c|c|c|c|c|c|c|c|c|c|c|c|c|c|c|c|c|c|c|}
\hline
$l$&1&2 &3 &4 &5  &6 &7 &8 &9 &10 &11&12 &13 &14 &15  &16 &17 &18 &19 &20 \\
\hline
$\alpha_l(\times 10^{-3})$&750&      563 &     422     & 316    &  237    &   179    &   134   &   100    & 75.2  &   56.2   &  42.6&32.0 &    23.8     & 17.9  &  13.4  &   10.1   & 7.54   & 5.70   & 4.31  &  3.19 \\
\hline
$\beta_l(\times 10^{-3})$&615&      382     & 239  &    151 &96.2&   62.1 &  40.2  & 26.0  & 17.1  & 11.2 &7.57 &  5.06  & 3.31 &  2.25 &  1.53 &  1.03  & 0.705  & 0.503   &0.335 &  0.230 \\
\hline
\end{tabular}
 \label{TableMoment}
 \vspace{-1em}
\end{table*}

\section{Numerical Results}\label{SectionSimulation}

\begin{figure}
\centering
   \includegraphics[width=.95\linewidth,  trim=65 0 80 15,clip]{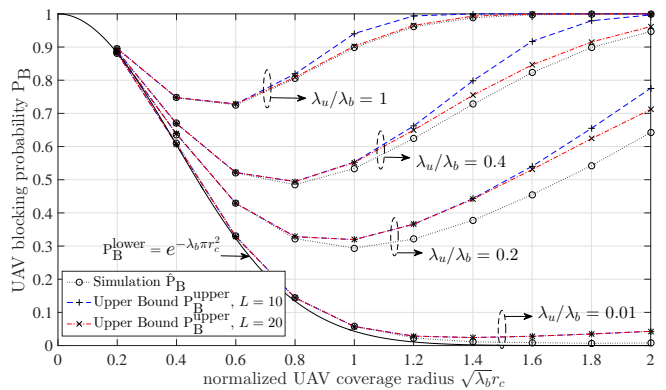}
\caption{UAV blocking probability $\textrm{P}_{\textrm{B}}$ versus normalized UAV coverage radius $\sqrt{\lambda_b}r_c$ under different UAV to BS density ratio $\lambda_u/\lambda_b$.\vspace{-2ex}}\label{rcCurve}
\end{figure}

First, we characterize the UAV blocking probability $\textrm{P}_{\textrm{B}}$ via simulations. 
Consider a HPPP $\Lambda_b$ with unit density $\lambda_b=1$ for BS locations and another independent HPPP $\Lambda_u$ for UAV locations with density $\lambda_u$ satisfying $\lambda_u/\lambda_b\leq 1$.
In the simulations, we generate the HPPPs $\Lambda_b$ and $\Lambda_u$ in a $100/\sqrt{\lambda_b}$-by-$100/\sqrt{\lambda_b}$ square region and obtain the average blocking probability $\hat{\textrm{P}}_{\textrm{B}}$ over 50 realizations under different $r_c$.
The simulation results are plotted in Fig. \ref{rcCurve} for different $\lambda_u/\lambda_b$.
We observe that the UAV blocking probability first decreases from 1 and then increases to 1 as $r_c$ increases from 0 to infinity, explained as follows. On one hand, for small $r_c$, increasing $r_c$ leads to larger UAV coverage area and hence is more likely to cover a BS.
In the special case with only one UAV in the network, the blocking probability decreases monotonically with $r_c$, and provides a lower bound as
\begin{equation}\label{PbLower}
\textrm{P}_{\textrm{B}}^{\textrm{lower}}=\exp(-\lambda_b\pi r_c^2)
\end{equation}
for the general case with multiple UAVs, which is also plotted in Fig. \ref{rcCurve}.
On the other hand, for large $r_c$, increasing $r_c$ results in more overlaps with other UAVs and hence less chance for a UAV to have an exclusively covered BS. In the limiting case with $r_c\rightarrow \infty$, all UAVs have zero residue non-overlapping coverage areas, and hence the blocking probability goes to 1.
It is also observed that $\textrm{P}_{\textrm{B}}$ is an increasing function of the UAV to BS density ratio $\lambda_u/\lambda_b$, which is expected since more UAVs lead to more overlaps and hence higher blocking probability.

\begin{figure}
\centering
   \includegraphics[width=1\linewidth,  trim=20 0 30 5,clip]{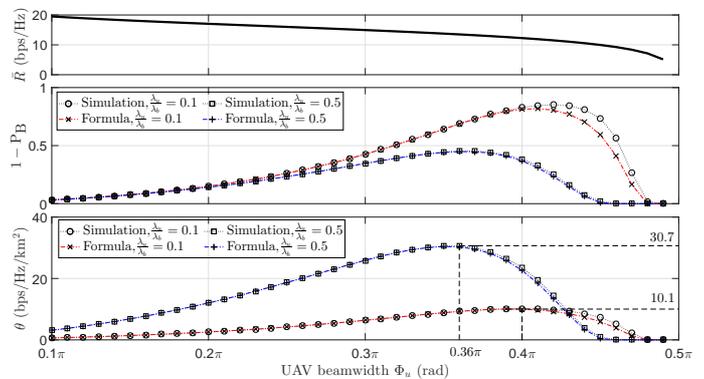}
\caption{Achievable rate $\bar R$, non-blocking probability $1-\textrm{P}_{\textrm{B}}$ and spatial throughput $\theta$ under different UAV beamwidth $\Phi_u$, given UAV altitude $H=100$ m and BS density $\lambda_b=10$ BSs/km$^2$.\vspace{-2ex}}\label{ThreeCurvePhiU}
\end{figure}

Next, we examine the tightness of our proposed upper bound in \eqref{LyuUpperBound} for the UAV blocking probability $\textrm{P}_{\textrm{B}}$. 
Note that in \eqref{LyuUpperBound}, we first need to obtain the estimated values for $\alpha_l$'s and $\beta_l$'s based on the method in Appendix \ref{appendixA}, where the results are summarized in TABLE \ref{TableMoment}.
For comparison, we plot in Fig. \ref{rcCurve} the proposed upper bound in \eqref{LyuUpperBound} with $L=10$ and $L=20$, respectively.
From the derivation of \eqref{Pout2termsUpper}, a larger $L$ provides a tighter upper bound for $\textrm{P}_{\textrm{B}}$ in general, as can be seen from Fig. \ref{rcCurve}.
We also simulated with $L$ values larger than 20, but only marginal improvement for the tightness is observed.
For $L=20$, it can be seen that the upper bound is tight for relatively large values of the UAV to BS density ratio, e.g., $\lambda_u/\lambda_b\geq 0.4$ in Fig. \ref{rcCurve}.
Moreover, for any ratio $\lambda_u/\lambda_b\leq 1$, the upper bound is tight in the region where $\textrm{P}_{\textrm{B}}$ monotonically decreases with $r_c$. This property makes the proposed spatial throughput lower bound in \eqref{throughputLower} more practically tight, since the maximum spatial throughput typically occurs in this region of the UAV blocking probability, as will be shown later (see Fig. \ref{ThreeCurvePhiU}).

Besides, we investigate the spatial throughput under various settings. 
The following parameters are used if not mentioned otherwise: BS density $\lambda_b=10$ BSs/km$^2$, $f_c=2$ GHz, $W=50$ kHz, $N_0=-174$ dBm/Hz and $P=0.5$ W.
Given the UAV altitude $H=100$ m, we plot in Fig. \ref{ThreeCurvePhiU} the achievable rate $\bar R$ in \eqref{RateUAVmin}, non-blocking probability $1-\textrm{P}_{\textrm{B}}$ and spatial throughput $\theta$ versus the UAV beamwidth $\Phi_u$ for different UAV to BS density ratio $\lambda_u/\lambda_b$, where $\textrm{P}_{\textrm{B}}$ is obtained by the proposed upper bound or by simulations.
It can be seen that $\bar R$ is a decreasing function of $\Phi_u$ with given $H$, while the non-blocking probability $1-\textrm{P}_{\textrm{B}}$ first increases and then decreases with $\Phi_u$.
As a result, the spatial throughput $\theta=\lambda_u(1-\textrm{P}_{\textrm{B}})\bar R$ first increases and then decreases with $\Phi_u$ for given $H$, $\lambda_b$, and $\lambda_u$.
Note that given $H$, smaller UAV beamwidth corresponds to smaller coverage area, which is less likely to cover any BS and thus results in lower non-blocking probability and lower spatial throughput. 
On the other hand, larger UAV beamwidth corresponds to larger coverage area, which is likely to cover more BSs but also results in more coverage area overlaps with other UAVs, thus rendering lower non-blocking probability and lower spatial throughput too. 
As a result,
for the case of $\lambda_u/\lambda_b=0.1$, the maximum spatial throughput $\hat\theta^*=10.1$ bps/Hz/km$^2$ is achieved at the optimal beamwidth $\Phi_u^*=0.4\pi$, while $\hat\theta^*=30.7$ bps/Hz/km$^2$ is achieved at $\Phi_u^*=0.36\pi$ for the case of $\lambda_u/\lambda_b=0.5$.
Note that the optimal UAV beamwidth $\Phi_u^*$ decreases as the UAV density increases with fixed BS density, so that the coverage radius $r_c=H\tan \Phi_u$ is reduced in order to avoid coverage area overlapping with more UAVs.
Similar performance trade-offs are observed under different UAV altitude $H$ with a fixed $\Phi_u$, where the results are omitted due to the space limitation.

Finally, we examine the tightness of our proposed lower bound \eqref{throughputLower} for the spatial throughput, which is also shown in Fig. \ref{ThreeCurvePhiU}.
It can be seen that our proposed achievable spatial throughput lower bound is tight for both cases of $\lambda_u/\lambda_b=0.1$ and 0.5, especially in the regime where the spatial throughput monotonically increases with the UAV beamwidth $\Phi_u$ (or equivalently where the UAV blocking probability decreases with $r_c$ given fixed $H$ as shown in Fig. \ref{rcCurve}).


In summary, our proposed upper bound in \eqref{LyuUpperBound} for the UAV blocking probability is practically tight, which also results in a tight lower bound in \eqref{throughputLower} for the achievable spatial throughput.
Moreover, adaptively adjusting the UAV altitude and/or beamwidth based on the BS/UAV densities helps to reduce the UAV blocking probability and hence can significantly improve the spatial throughput of the cellular-enabled UAV network.

\section{Conclusions}\label{SectionConclusion}

In this paper, we study a new cellular-enabled UAV communication network and propose the use of a directional antenna at each UAV to limit the interference to/from other UAVs, and hence improve the performance of spatial frequency reuse. We first derive a closed-form upper bound for the UAV communication blocking probability and then characterize the achievable average spatial throughput of all UAV users in the network. 
Via extensive simulations, the derived bounds are verified to be practically tight. Moreover, it is shown that the spatial throughput can be significantly improved by optimizing the UAV altitude and/or beamwidth under different BS/UAV densities.   

We have so far considered a single layer of UAV users at a common altitude, which can be extended in future work to the more general scenario with UAVs at different altitude.
Moreover, in this paper we have considered a single channel to be shared by UAVs exclusively, while it is worth investigating the general case of multi-channels shared by UAVs as well as ground users.

\appendices
\section{Method to Obtain/Estimate $\alpha_l$'s and $\beta_l$'s}\label{appendixA}
For the special case with $l=0$, we have $\eta=1$ and hence $\alpha_0=1, \beta_0=1$. Therefore, from \eqref{Ketal} we have
\begin{equation}\label{term0}
\textrm{Pr}\big\{\mathcal{W}\cap \mathcal{A}_0=\emptyset\big | |\mathcal{M}|=0\big\}=\exp(-\lambda_b S)=K_{\eta,0}(-\lambda_b S).
\end{equation}
For $l=1$, there is only one UAV $m$ overlapping with UAV 0, i.e., UAV $m$ is at a distance $r\triangleq \|\bold u_m-\bold u_0\|\leq 2 r_c$ from UAV 0.
From geometry, the overlapping area of two disks of radius $r_c$ at distance $r$ is given by 
\begin{equation}
C(r)=2r_c^2\arccos(\frac{r}{2 r_c})-r\sqrt{r_c^2-\frac{r^2}{4}}, \quad 0\leq r\leq 2 r_c.
\end{equation}
Since UAV $m$ is uniformly distributed within distance $r\leq 2 r_c$ from UAV 0, the average non-overlapping coverage area $\mathbb{E}\big[A\big | |\mathcal{M}|=1\big]$ is given by
\begin{small}
\begin{align}
&\mathbb{E}\big[A\big | |\mathcal{M}|=1\big] =\int_{A=0}^S A f\big(A\big | |\mathcal{M}|=1\big)\diff A\notag\\
&=\int_{\phi=0}^{2 \pi}\int_{r=0}^{2 r_c} \big(\pi r_c^2-C(r)\big) \frac{1}{\pi(2 r_c)^2}r\diff r \diff \phi=\frac{3}{4}\pi r_c^2=\frac{3}{4}S.
\end{align}
\end{small}
Similarly, we can obtain
\begin{equation}
\mathbb{E}\big[A^2\big | |\mathcal{M}|=1\big]=\big(\frac{3}{4}-\frac{4}{3\pi^2}\big)S^2.
\end{equation}
Therefore, we have $\alpha_1=\frac{3}{4}$ and $\beta_1=\frac{3}{4}-\frac{4}{3\pi^2}\approx0.615$.
For the cases with $l\geq 2$, the non-overlapping coverage area $A$ is complicated to characterize. Fortunately, we can estimate 
the $\alpha_l$'s and $\beta_l$'s via computer simulations.

We resort to Monte-Carlo simulations to estimate $\eta$ and then $\alpha_l$'s and $\beta_l$'s.
Specifically, for a given $l\geq 2$, we can generate a topology with $l$ UAVs independently and uniformly located in the disk region $\mathcal{B}(\bold u_0,2 r_c)$ with radius $2r_c$ centered at the ground projection $\bold u_0$ of UAV 0 (see, e.g., the case of $l=2$ in Fig. \ref{Cellular4UAVa0}(c)).
For each topology, we can generate $Q=10^{5}$ points uniformly distributed in the coverage disk $\mathcal{B}(\bold u_0,r_c)$ of UAV 0 and count the number of points $q$ inside the non-overlapping coverage region $\mathcal{A}_0$, from which the non-overlapping ratio $\eta$ can be estimated as $\hat\eta=q/Q$.
In this way, we can independently generate $T=10^{5}$ topologies for a given $l$ and obtain an $\hat\eta$ for each topology. The $\alpha_l$'s and $\beta_l$'s can then be estimated from the $T$ samples of $\hat\eta$ for a given $l$, where the results are summarized in TABLE \ref{TableMoment}.

\bibliography{IEEEabrv,BibDIRP}

\begin{thebibliography}{10}
\providecommand{\url}[1]{#1}
\csname url@samestyle\endcsname
\providecommand{\newblock}{\relax}
\providecommand{\bibinfo}[2]{#2}
\providecommand{\BIBentrySTDinterwordspacing}{\spaceskip=0pt\relax}
\providecommand{\BIBentryALTinterwordstretchfactor}{4}
\providecommand{\BIBentryALTinterwordspacing}{\spaceskip=\fontdimen2\font plus
\BIBentryALTinterwordstretchfactor\fontdimen3\font minus
  \fontdimen4\font\relax}
\providecommand{\BIBforeignlanguage}[2]{{%
\expandafter\ifx\csname l@#1\endcsname\relax
\typeout{** WARNING: IEEEtran.bst: No hyphenation pattern has been}%
\typeout{** loaded for the language `#1'. Using the pattern for}%
\typeout{** the default language instead.}%
\else
\language=\csname l@#1\endcsname
\fi
#2}}
\providecommand{\BIBdecl}{\relax}
\BIBdecl

\bibitem{UAVinterferenceMagazine}
B.~V.~D. Bergh, A.~Chiumento, and S.~Pollin, ``{LTE} in the sky: trading off
  propagation benefits with interference costs for aerial nodes,'' \emph{IEEE
  Commun. Mag.}, vol.~54, no.~5, pp. 44--50, May 2016.

\bibitem{SkyIsNotLimit}
\BIBentryALTinterwordspacing
X.~Lin \emph{et~al.}, ``The sky is not the limit: {LTE} for unmanned aerial
  vehicles.'' [Online]. Available: \url{http://arxiv.org/abs/1707.07534}
\BIBentrySTDinterwordspacing

\bibitem{3GPPworkItemUAV}
RP-170779, ``{New SID on Enhanced Support for Aerial Vehicles},'' in \emph{3GPP
  TSG RAN Meeting 75}.\hskip 1em plus 0.5em minus 0.4em\relax NTT DOCOMO INC,
  Ericsson, Mar. 2017.

\bibitem{KDDItrials}
R1-1705823, ``{Field measurement results for drone LTE enhancement},'' in
  \emph{3GPP TSG-RAN WG1 Meeting 88bis}.\hskip 1em plus 0.5em minus 0.4em\relax
  KDDI Corporation, Mar. 2017.

\bibitem{QualCommDroneReport}
Qualcomm, \emph{{LTE Unmanned Aircraft Systems---Trial Report}}.\hskip 1em plus
  0.5em minus 0.4em\relax Qualcomm Technologies, Inc., May 2017.

\bibitem{ZengUAVmag}
Y.~Zeng, R.~Zhang, and T.~J. Lim, ``Wireless communications with unmanned
  aerial vehicles: opportunities and challenges,'' \emph{IEEE Commun. Mag.},
  vol.~54, no.~5, pp. 36--42, May 2016.

\bibitem{CyclicalLyu}
J.~Lyu, Y.~Zeng, and R.~Zhang, ``Cyclical multiple access in {UAV}-aided
  communications: a throughput-delay tradeoff,'' \emph{IEEE Wireless Commun.
  Lett.}, vol.~5, no.~6, pp. 600--603, Dec. 2016.

\bibitem{PlacementLyu}
J.~Lyu, Y.~Zeng, R.~Zhang, and T.~J. Lim, ``Placement optimization of
  {UAV}-mounted mobile base stations,'' \emph{IEEE Commun. Lett.}, vol.~21,
  no.~3, pp. 604--607, Mar. 2017.

\bibitem{UAVGBSglobecom}
\BIBentryALTinterwordspacing
J.~Lyu, Y.~Zeng, and R.~Zhang, ``Spectrum sharing and cyclical multiple access
  in {UAV}-aided cellular offloading,'' in \emph{Proc. IEEE GLOBECOM 2017}, to
  appear. [Online]. Available: \url{https://arxiv.org/abs/1705.09024}
\BIBentrySTDinterwordspacing

\bibitem{DroneSmallCell}
M.~Mozaffari, W.~Saad, M.~Bennis, and M.~Debbah, ``Drone small cells in the
  clouds: Design, deployment and performance analysis,'' in \emph{Proc. IEEE
  GLOBECOM}, Dec. 2015, pp. 1--6.

\bibitem{3Dplacement}
R.~I. Bor-Yaliniz, A.~El-Keyi, and H.~Yanikomeroglu, ``Efficient {3-D}
  placement of an aerial base station in next generation cellular networks,''
  in \emph{Proc. IEEE Int. Conf. Commun. (ICC)}, May 2016, pp. 1--5.

\bibitem{ZhangWeiSmallCellJSAC}
C.~Zhang and W.~Zhang, ``Spectrum sharing for drone networks,'' \emph{IEEE J.
  Sel. Areas Commun.}, vol.~35, no.~1, pp. 136--144, Jan. 2017.

\bibitem{UAVoverload}
S.~Rohde and C.~Wietfeld, ``Interference aware positioning of aerial relays for
  cell overload and outage compensation,'' in \emph{Proc. IEEE Vehicular
  Technology Conference (VTC)}, Sep 2012, pp. 1--5.

\bibitem{UAVpublicSafety}
A.~Merwaday and I.~Guvenc, ``{UAV} assisted heterogeneous networks for public
  safety communications,'' in \emph{Proc. IEEE Wireless Commun. Netw. Conf.
  (WCNC)}, Mar. 2015, pp. 329--334.

\bibitem{WuQingQingUAVTWC}
\BIBentryALTinterwordspacing
Q.~Wu, Y.~Zeng, and R.~Zhang, ``Joint trajectory and communication design for
  multi-{UAV} enabled wireless networks.'' [Online]. Available:
  \url{https://arxiv.org/abs/1705.02723}
\BIBentrySTDinterwordspacing

\bibitem{UAVrelay}
P.~Zhan, K.~Yu, and A.~L. Swindlehurst, ``Wireless relay communications with
  unmanned aerial vehicles: Performance and optimization,'' \emph{IEEE Trans.
  Aerosp. Electron. Syst.}, vol.~47, no.~3, pp. 2068--2085, July 2011.

\bibitem{ZengMobileRelay}
Y.~Zeng, R.~Zhang, and T.~J. Lim, ``Throughput maximization for {UAV}-enabled
  mobile relaying systems,'' \emph{IEEE Trans. Commun.}, vol.~64, no.~12, pp.
  4983--4996, Dec. 2016.

\bibitem{ZhanChengDataCollection}
\BIBentryALTinterwordspacing
C.~Zhan, Y.~Zeng, and R.~Zhang, ``Energy-efficient data collection in {UAV}
  enabled wireless sensor network.'' [Online]. Available:
  \url{https://arxiv.org/abs/1708.00221}
\BIBentrySTDinterwordspacing

\bibitem{balanis2016antenna}
C.~A. Balanis, \emph{Antenna theory: analysis and design}.\hskip 1em plus 0.5em
  minus 0.4em\relax John Wiley \& Sons, 2016.

\bibitem{DBrook}
D.~Brook, ``Bounds for moment generating functions and for extinction
  probabilities,'' \emph{Journal of Applied Probability}, vol.~3, no.~1, pp.
  171--178, 1966.

\end{thebibliography}

\newpage

\end{document}